\documentclass{ws-procs9x6}

\begin{document}

\title{Quark Model Perspectives on Pentaquark Exotics}

\author{K. Maltman}

\address{Dept. Math. and Stats., York Univ., 4700 Keele St. \\
Toronto, ON CANADA M3J 1P3\\ 
E-mail: kmaltman@yorku.ca}
\address{CSSM, University of Adelaide, Adelaide, SA 5005 AUSTRALIA}

\maketitle

\abstracts{I discuss the expectations and predictions for pentaquark exotics
based on the quark model perspective. Recent quark model scenarios,
and calculations performed in different realizations of the
quark model approach, up to the end of March 2004, are also
discussed.}

\section{Introduction}
A large number of experiments now appear to confirm the 
existence of the exotic, strangeness $+1$ $\theta$ baryon\cite{thetaexp}. 
The $I=3/2$ exotic $\Xi_{3/2}$ signal observed by NA49
remains to be confirmed (it is not seen by HERA-B,
and its compatibility with earlier
high statistics $\Xi$ production experiments has also been
questioned)\cite{xi32}. Recently, H1 reported evidence
for an anti-charmed exotic, though this state was not 
seen by ZEUS\cite{H1zeus}.

Whether or not the NA49 and H1 signals are confirmed by subsequent
experiments, the existence of the $\theta$ makes
a rethinking of our understanding of the excited baryon spectrum
inevitable. If the $\theta$ has $I=0$, and lies in a
$\overline{10}_F$ multiplet, for example, 
exotic pentaquark partners having $N$ and $\Sigma$ quantum numbers
necessarily also exist. These should sit 
in the same region as the $3q$ radial excitations of
the $N$ and $\Sigma$ ground states and, unavoidably,
mix with them{\footnote{Mixing
between non-exotic radial excitations and the
corresponding states in the exotic $\overline{10}_F$ and
$27_F$ multiplets is also significant
in the chiral soliton model\cite{weigelnew}. In
the quark model, where additional non-exotic pentaquark states 
are expected, the mixing will be even more complicated.
Phenomenologically, a more complicated mixing 
pattern than just ideal mixing between $\overline{10}_F$
and $8_F$ pentaquark multiplets\cite{jw} is likely required to 
account for the $N(1440)$ and $N(1710)$ masses and decay 
patterns\cite{cohensuzuki}.}}.
This immediately calls into question past quark model
treatments of the excited, positive parity baryon spectrum which 
included only $3q$ configurations. It also undercuts 
one of the main phenomenological motivations for
the effective Goldstone boson (GB) exchange model of the
baryon spectrum\cite{glozman}, i.e., the failure of the $3q$
Isgur-Karl (effective color-magnetic (CM) exchange) approach to
successfully reproduce the low-lying $P=+$ Roper-like resonances.
The existence of the $\theta$ does not, of course, 
invalidate the GB model, but does suggest that any differences between
GB and CM model predictions in the exotic sector become of
heightened phenomenological interest.

Below, we discuss recent scenarios, and some qualitative features
of pentaquark states expected in the quark model (QM) framework.
Comparisons to the results of the 
chiral soliton model (CSM) approach\cite{earlycsm,dpp,dppothers,ekp}, 
whose prediction of a low-lying, {\it narrow} $\theta$\cite{dpp} 
was a primary motivation for the initial LEPS search, will
also be made.

\section{The $\theta$ Parity and Other Discrete Quantum Numbers}
The CSM approach unambiguously predicts that the lowest lying
$S=+1$ exotic state should lie in the $\overline{10}_F$ multiplet
and have $I=0$, $J^P=1/2^+$. It has sometimes been
stated that the naive quark model ``predicts'' $P=-$ for the lowest lying
exotic baryon state. This statement is incorrect
and has led to some confusion
in the literature. It should actually be rephrased to state that
the quark model ``might naively be guessed to produce $P=-$''
for the lowest-lying exotic state. Whether or not this guess is 
correct is a dynamical question. In fact, it turns out that
a competition exists between the additional orbital excitation 
needed for the $P=+$ sector and the decreased spin-dependent (generically,
``hyperfine'') expectation available in this sector. 

The following qualitative argument, given by Jaffe and Wilczek 
(JW)\cite{jw,nuss}, shows why the $P=+$ sector might be favored.
The $F=\bar{3}$, $J=0$, $C=\bar{3}$ $qq$ configuration
is known to be very attractive in QCD. It is also the most
attractive $qq$ correlation in a number of QCD-inspired models
(the GB and CM models, as well as models based on 
instanton-induced effective interactions). Assuming 
pentaquark states are dominated by optimal two-quark correlations,
one expects a state with two such pairs to be particularly low-lying.
Such a state is Pauli forbidden unless the two pairs 
are in an odd relative orbital state\cite{jw}. To take advantage of 
this optimal $qq$ pairing, one must thus go to the $P=+$ sector. 
The lowest-lying exotic configuration is
then necessarily the $S=+$, $I=0$ member of a $\overline{10}_F$, $J^P=1/2^+$
multiplet, as in the CSM picture.

A qualitative understanding of why the hyperfine energy
might be significantly lowered in the $P=+$ 
sector, and hence win out over the orbital excitation, can
can also be arrived at using the ``schematic approximation'' to 
the GB and CM models, in which the spatial
dependence of the spin-dependent operators is neglected{\footnote{The
approximation has also been employed quantitatively
in a number of recent calculations\cite{schematics}. However, although
it successfully identifies optimally attractive channels,
it turns out to be {\it quantitatively} unreliable (see Ref. [14]
for more details).}}.
In this approximation, the expectations of the flavor-spin (FS) (GB case) 
or color-spin (CS) (CM case) dependent interactions can be worked 
out by group-theoretic methods, even in the $P=+$ sector.
In both models, higher FS or CS symmetries produce more attractive 
hyperfine expectations. For the GB case, the highest FS symmetry for the 
spatially-unexcited $[4]_L$ orbital $q^4$ configuration 
is $[31]_{FS}$, while for the $[31]_L$ configuration it is $[4]_{FS}$.
Similarly, for the multiplets containing exotic states in the CM case,
the highest $CS$ symmetries are $[22]_{CS}$ for the $[4]_L$
and $[31]_{CS}$ for the $[31]_L$ configuration.
In both cases one thus expects a significant gain in hyperfine
energy in going from the $P=-$ to the $P=+$ sector. 
Explict dynamical model calculations bear this out\cite{jm}.

Dynamically, it need not be the case that $qq$
correlations are dominant. Indeed, in the CM model,
as pointed out by Karliner and Lipkin (KL)\cite{kl}, 
a more complicated correlation, consisting of one 
$F=\bar{3}$, $C=\bar{3}$, $J=0$ pair (as in the JW scenario)
and one $F=\bar{3}$, $C=\bar{3}$ pair with the $qq$ spin 
flipped to $J=1$ and anti-aligned to the $\bar{s}$ spin,
yields a lower hyperfine energy for the $\theta$
than does the JW correlation.
(Such a configuration is also favored 
in a model with effective instanton-induced interations\cite{vento}.)
Mixing between the JW and KL correlations in the CM model, induced
by the same $q\bar{s}$ interactions responsible for favoring
the KL $qq\bar{s}$ correlation, actually leads to an even
lower-lying state, which is nearly an equal mixture of the
JW and KL correlations\cite{jm}.

It should be stressed that, while in the JW and KL scenarios it
has been argued that intercluster interactions and antisymmetrization 
effects will be suppressed by the relative $p$-wave between the 
clusters, it is only in particular dynamical models 
that these effects can be explicitly calculated.
Such a calculation was performed for the GB and CM models in Ref. [14].
In such calculations, one can directly compare
the hyperfine expectations in the $P=-$
and $P=+$ sectors. As shown in Ref. [14], at least for
the GB and CM models, the increase in hyperfine attraction in
the optimal (CSM quantum numbers) $P=+$ channel, as compared to
that in the optimal non-fall-apart $P=-$ channel, is such that,
with expectations for the orbital excitation energy based on
experience from the baryon sector\cite{kl}, the lowest-lying
exotic state is expected to have $P=+$ and {\it NOT} $P=-$, 
with other quantum numbers also agreeing with the CSM 
prediction{\footnote{An even stronger statement is true
in the CM model. There, even if one argues that the approach
used in Ref. [15] might mis-estimate the orbital excitation energy,
the model allows no phenomenologically acceptable $P=-$ assignment for
the $\theta$\cite{jm}.}}.

Two important qualitative differences do exist between the CSM and
QM pictures. The first difference concerns ``flavor partners''.
In the QM picture, in the absence of flavor-dependent $q\bar{q}$
interactions, the
exotic flavor multiplets come accompanied by non-exotic flavor
partners with which they are degenerate in the $SU(3)_F$ limit.
For example, the $4q$ flavor configuration in the $\overline{10}_F$
multiplet is $[22]_F$. Combining this with the $[11]_F$
$\bar{q}$ configuration yields
\begin{equation}
[22]_F\otimes [11]_F\, =\,  {\overline{10}_F}\oplus 8_F\ ,
\label{fpartner}\end{equation}
i.e., the $\overline{10}_F$ pentaquark multiplet containing the $\theta$
is accompanied by an $8_F$ pentaquark multiplet{\footnote{The
exotic $27_F$ pentaquark multiplet similarly comes accompanied
by a $10_F$ and an $8_F$, the $35_F$ pentaquark multiplet by
a $10_F$.}}. When $SU(3)_F$
breaking is turned on, the $N$ and $\Sigma$ partners
of the $\theta$, the members of the pentaquark $8_F$, and the
radially excited $3q$ configurations will all mix. Thus, if the
$\theta$ is, indeed, real, the $P=+$ excited baryon sector 
becomes very complicated in the QM picture. 
The second difference between the QM and CSM pictures
is that $P=+$ pentaquarks in the QM approach
are accompanied by spin-orbit partners not present in the CSM.
For the $\theta$, for example,
the intrinsic spin of $1/2$, coupled to the $L=1$ of the
orbital excitation in the $P=+$ sector, leads to both
$J^P=1/2^+$ and $3/2^+$ $S=+$, $I=0$ states. While a low-lying
$S=+$, $J^P=3/2^+$ state {\it is} predicted in the CSM approach,
it lies in a $27_F$, and has $I=1$, not $I=0$. An estimate
of the expected splitting of the $J^P=3/2^+$ partner of
the $\theta$ in the CM model suggests it should be rather
small, $\sim$ several $10$'s of MeV, with
a conservative maximum of $150$ MeV\cite{cdls}. Such observations
make the importance of searches for excitations of the $\theta$
obvious.

\section{Masses of Exotic States}
The CSM approach naturally predicts a low-lying $S=+$ exotic
with a mass in the region of the observed experimental $\theta$
signal (see Ref. [11] for a detailed discussion of this point). In contrast,
simple extensions of constituent quark model calculations from the
non-exotic $3q$ baryon sector to the exotic sector will 
produce a mass for the lowest such exotic which is too high.

It is important to bear in mind that, although it is 
not unreasonable to attempt such calculations as an exploratory
first stage, there are good reasons
for expecting them to be physically unreliable, even if the underlying models
on which they are based are reasonable. The reason is that the models
typically lack a representation of physical effects which one
expects to be present and to, potentially, have a significant
impact on the values of one-body energies. An example of such
effects is provided by the bag model. In going from the $3q$ to
$6q$ sector, for example, the equilibrium bag radius increases, reducing
the quark kinetic energies. This effect is counterbalanced by
the change in the phenenological $Z/R$ term, meant to represent
the effects of zero point motion and corrections for CM motion
in the bag. It turns out that each of these changes is
large ($\sim 400-450$ MeV) on the scale of baryon splittings,
and that the level of cancellation between them is a {\it very} sensitive
function of the bag parameter $B$\cite{kmdib}. Such effects
are almost certainly present physically, and in need of representation
if one wants to generalize calculations from the $3q$ to
the $4q\bar{q}$ sector. They are not, however, represented at all
in constituent quark model approaches such as those of the
GM and CM models. As a result, one would not generally
expect the one-body energies, calculated in those versions of the 
models calibrated in the $3q$ sector, to be reliable in the
pentaquark sector. It thus appears fair to say both that the 
$\theta$ mass has {\it not} been predicted, and that
it most likely {\it cannot} be sensibly predicted,
in the QM framework.

This does {\it not} mean that the various quark models
cannot make any predictions in the pentaquark sector, only
that, realistically, they lack the features required to allow them
to have a chance of successfuly predicting the splitting between (exotic)
pentaquark and (non-exotic) $3q$ states. For example, 
one of the assumptions of the 
models is that the spin-dependent interactions can, to a good
approximation, be treated perturbatively. If this is the case,
then the splittings between different spin-flavor channels,
all within the pentaquark sector, should still be predictable
by the models. Failure of experiment to reproduce these
splittings would then allow one to rule out a given model, or models.

The minimal model-dependence for such predicted splittings occurs
for $4q\bar{q}^\prime$ states where all of the four quarks are
$u$ and/or $d$. When there are both $u$ (or $d$) and $s$ quarks
among the four quarks, there can be a model-dependent interplay
between the flavor-breaking in the hyperfine expectations and the lowering of
orbital excitation energies for relative coordinates involving
the heavier $s$ quark(s). One of the interesting predictions,
of the minimally-model-dependent type, is that, as in the CSM,
a rather low-lying $I=1$, $S=+$ excitation, $\theta_1$, of the $\theta$
should exist in both the GB and CM models. 
In the GB model there is actually a degenerate pair with
$(I,J^P_q)=(1,1/2^+)$ and $(1,3/2^+)$, where $J_q$ is
the total quark spin (still to be combined with the orbital
$L=1$ to produce the total spin). In the CM model, the
lowest excitation of the $\theta$ has $(I,J^P_q)=(1,1/2^+)$.
Using non-exotic baryon values of
the pair hyperfine matrix elements to estimate the hyperfine
energies one finds\cite{jm}
\begin{eqnarray}
&&m_{\theta_1}-m_\theta \simeq 60-90\ {\rm MeV}\ {\rm (CM)}\nonumber\\
&&m_{\theta_1}-m_\theta \simeq 140\ {\rm MeV}\ {\rm (GB)}\ ,
\label{theta1mass}\end{eqnarray}
to be compared to $\simeq 55-85$ MeV in the rigid rotor version
of the CSM approach\cite{dppothers,ekp}.

Estimates for the splitting between the $\theta$ and its $I=3/2$
$\Xi_{3/2}$ $\overline{10}_F$ partner have been made in
both the JW and KL scenarios. Both the original version of the
JW estimate and the KL estimate, which yielded $m_{\Xi_{3/2}}\simeq 1750$ MeV
and $\simeq 1720$ MeV, respectively, were based on the assumption
that the pair matrix elements for the spin-dependent interactions,
and the cost of the replacement $d\leftrightarrow s$, could
be estimated using the analogous quantities from the non-exotic 
baryon sector. The JW estimate can be raised to $\sim 1850$, more
in line with the NA49 observation, if one allows significant
deviations from the non-exotic baryon sector parameter values\cite{jw2}.
One should again bear in mind that cross-cluster interaction and
antisymmetrization effects, where novel flavor-breaking contributions
might be generated, are implicitly neglected in these estimates.
More detailed dynamical model estimates will be subject to
the model-dependence noted above, associated with the need to estimate
flavor-breaking effects on the one-body energies. 
One such dynamical calculation has been performed, for the GB
model, in Ref. [20], with the result
$m_{\Xi_{3/2}}\simeq 1960$ MeV. Note, however, that, while the
actual calculation is non-schematic, the wavefunction is
restricted to the single component which lies lowest in
the schematic approximation. While, for technical reasons
having to do with the explicit form of the flavor-spin
interactions employed in the model, this approximation
is a good one for the $S=+$ sector (where all four quarks
have equal mass), there are reasons
to expect much more significant mixing in the $\Xi_{3/2}$ sector
once the schematic approximation is relaxed.
Allowing additional components in the wavefunction 
will lower the mass. The size of this effect is not known
at present.

The lowest $I=2$ $S=+$ exotics, using non-exotic baryon
values for the two-body spin-dependent matrix elements, are predicted to 
lie around $\sim 1980$ MeV in both the GB and CM models, 
similar to the values obtained in
the CSM approach. Both experiment and theory, therefore, strongly
disfavor an $I=2$ interpretation of the $\theta$.

\section{The $\theta$ Width}
One of the striking predictions of the CSM calculation of
Ref. [9] was that the $\theta$ should be naturally
narrow ($\sim$ a few $10$'s of MeV or less) in the CSM
picture, as subsequently observed experimentally.
Some initial speculations, based on the observed widths
of known, non-exotic baryons a comparable distance above
their own two-body decay thresholds, suggested that
the $\theta$ should be relatively broad in the QM
picture. Such arguments, however, are necessarily
unreliable since the decay mechanism for the non-exotic
$3q$ and exotic pentaquark baryons in the quark model cannot be the same.
Indeed, for two-body decays of a $3q$ baryon,
a pair creation is required whereas, for the decay of a pentaquark
state, the number of constituents is the same in the initial and final
states.

If one considers $KN$ scattering, and the possibility of forming an
$S=+$ exotic resonance as a result of the residual
short-range interaction among the fixed number of
constituents, one realizes that an above-threshold
resonance can only be formed in a $p$-wave or higher.
The reason, as stressed in Ref. [6], is that a single-range residual
$s$-wave interaction insufficiently strong to bind produces
no resonance behavior, only positive phase motion.
In contrast, for $p$-wave (and higher) scattering,
a residual short-range attraction can play off against the
peripheral centrifugal barrier to produce resonance behavior.
One can make a rough estimate for the width of such a
resonance as follows. It is straightforward to verify
that the intrinsic width for a $KN$ resonance at the
observed mass of the $\theta$ produced
by an attractive $KN$ square-well potential of hadronic size is
$\sim 200$ MeV\cite{jw}. Thus, if one has a pentaquark
configuration with overlap $f$ to the short-range $KN$
configuration, one expects a width, for a $p$-wave resonance,
of order
\begin{equation}
\Gamma_\theta \sim 200\, f^2\ {MeV}\ .
\label{width}\end{equation}
Whether or not the small widths compatible with experimental
observations are natural in the QM picture is then
a matter of how large or small the overlap factor $f$ is.

It turns out that, for the JW correlation, the isospin-spin-color
part of the overlap factor is rather small,
$\left[ f_{ISC}^{JW}\right]^2=1/24$\cite{jm}. 
A similar value, $\left[ f_{ISC}^{CM}\right]^2\simeq 1/25$,
is obtained for the optimized combination of the 
JW and KL correlations in the CM model\cite{jm}. Since these results 
do not include any further reduction associated with 
the mismatch between the spatial configurations (which can 
be numerically quite significant\cite{carlson2}),
the natural width of the $\theta$ in the QM picture
is, in fact, quite small. Indeed, a width greater than
$\sim 10$ MeV would be very difficult to accommodate.
$SU(3)$ arguments then require the width of the $\Xi_{3/2}$
partner of the $\theta$ to also be small\cite{jw,jw2}.

It is obvious that the above width estimate is at best
semi-quantitative. Unfortunately,
it seems unlikely that significant improvements can be made to 
it{\footnote{The discussion of Ref. [11]
shows that, because of cancellations between nominally
leading-order contributions to the widths of the $\overline{10}_F$
states, it is similarly difficult to provide a quantitatively
reliable prediction for these widths in the CSM approach. 
Such cancellations can also amplify the impact of higher
order $SU(3)_F$-breaking effects on the relation between
the $\theta$ and $\Xi_{3/2}$ widths.}}.
The most natural improvement one could envisage, in the
GB and CM models, would be to use the non-relativistic
constituent QM framework,
where CM motion can be cleanly separated, and do a scattering
calculation of the resonating group type.
The obvious difficulty
with this approach is that the one-body operators enter such
a calculation in a non-trivial fashion. The existence of
problems with the one-body energies in such models thus means that
resonance widths obtained in such a calculation could not be
treated as reliable.

Finally, it should be mentioned that a common coupling of nearby
states to the same decay channel can lead, through mixing, to
one of the mixed states having a width much narrower than the
natural width of either state\cite{couplingwidth}. To produce
a significant narrowing, the mechanism requires the two
states, before mixing, to be relatively close together. For
the GB and CM models, the next excitation with $\theta$
quantum numbers lies $\sim 330$ MeV (GB) and
$\sim 230$ MeV (CM) above the $\theta$\cite{jm}. In these
models, therefore, the mixing mechanism is unlikely to play a significant
role in generating the narrow $\theta$ width. This does not, of course,
preclude the possibility that the mixing mechanism might be important 
in other realizations of the QM approach.

\section{Heavy Quark Analogues of the $\theta$}
Interest in heavy pentaquarks ($\bar{Q}q^4$, 
where $q=u,d,s$ and $\bar{Q}=\bar{c},\bar{b}$), 
was initially aroused by the observation that the 
$\bar{Q}s\ell^3$ ($\ell =u,d$) states with $I=1/2$, $J^P=1/2^-$
have strong hyperfine attraction relative
to that of their two-body decay thresholds, $ND_s$, 
$NB_s$, in the CM model, in the $m_{\bar{Q}}\rightarrow\infty$
limit\cite{earlyheavy}.
Subsequent work, however, showed that decreased binding from $SU(3)_F$ 
breaking, kinetic energy, confinement and $m_{\bar{Q}}\not= \infty$ effects
was likely sufficient to make all of these states unbound\cite{nextheavy}.

Predictions turned out to be very
different in the GB model, with the $P=-$ states
lying several $100$ MeV above threshold\cite{gbnegparheavy}. Only
the $\bar{Q}\ell^4$ $P=+$, $(I,J_q^P)=(0,1/2^+)$ states
were found to be bound, with
binding energies of $75-95$ MeV\cite{gbposparheavy}.

An experimental search for the predicted anticharmed, 
strange state, covering the mass range $2.75-2.91$ GeV, was 
performed by the E791 Collaboration, with negative results\cite{e791heavy}.

Interest in heavy pentaquark states has been greatly revived
by the discovery of the $\theta$. If, as is now generally assumed,
the parity of the $\theta$ is indeed positive, then the same
mechanism which makes the $\theta$ narrow is expected to also
make its heavy quark analogues narrow, even if they lie above
the relevant nucleon-plus-heavy-pseudoscalar decay threshold.
The situation for the $P=-$ heavy pentaquarks is less clear.
Models, as well as a JW-like scenario for the $\bar{Q}s\ell^3$
states\cite{wiseetal}, suggest that the lowest-lying of these
states should have $J=1/2$. Unless such a state is bound,
it will have an $s$-wave fall-apart decay and hence almost certainly
be non-resonant. 

A number of recent estimates exist for the $P=+$ heavy pentaquark
masses\cite{jw,klthetac,huang}. These are typically produced by
extensions of the scenarios for the $\theta$ based on the
assumption that a reasonable approximation to the
splitting between the $\theta$ and its $I=0$, $J^P=1/2^+$ 
analogue, $\theta_c$ or $\theta_b$, should
be obtainable using the ``corresponding'' splitting
between the $\Lambda$ and $\Lambda_c$ or $\Lambda_b$, supplemented
by an estimate for the change (if any) in the spin-dependent
quark-antiquark interactions in going from the $\theta$ to
the heavy quark system. Since, 
in the JW scenario, the diquarks are assumed to have spin zero
and be tightly bound, there is no such quark-antiquark
interaction, and hence no spin-dependent correction to
be made. The resulting estimates are\cite{jw}
\begin{equation}
m_{\theta_c}\simeq 2710\ {\rm MeV};\quad 
m_{\theta_b}\simeq 6050\ {\rm MeV}\ ,
\label{jwthetac}\end{equation}
$\sim 100$ and $170$ MeV below the relevant strong decay thresholds.
If one assumes that the same diquark-triquark clustering 
postulated in the KL scenario for the $\theta$
persists for heavy systems, one obtains, after 
taking into account the reduced strength of the $\bar{Q}\ell$,
relative to $\bar{s}\ell$, hyperfine interaction,
\begin{equation}
m_{\theta_c}\simeq 2985\ {\rm MeV};\quad 
m_{\theta_b}\simeq 6400\ {\rm MeV}\ ,
\label{klheavy}\end{equation}
now $\simeq 180$ MeV above the relevant strong decay thresholds.
The estimate of Ref. [28] for the low-lying $\bar{Q}s\ell^3$, $P=-$
states is based on the JW scenario, and the JW estimate for the $P=+$ 
states. Estimates for the reduction in mass
associated with the absence of an orbital excitation, and
the increase in mass associated with changing one of the $u,d$
quarks of the $\theta_c$ to an $s$, both are taken from analogous
splittings in the ordinary charmed and charm-strange baryon spectrum. 
While the neglect of cross-cluster interactions and 
antisymmetrization effects is more questionable when the
diquark clusters are in a relative $s$-wave, the resulting
estimate is of interest since it puts the $\bar{Q}s\ell^3$ states
not only below strong decay thresholds (at $2580$ and
$5920$ MeV for $Q=c,b$, respectively) but also below the
lower edge of the E791 search window in the $Q=c$ case.

It should be pointed out that the KL assumption that the same diquark-triquark
clustering is present in both the heavy quark and $\theta$
systems requires some deviation from the strict CM model picture.
The reason is that the constituent charm quark mass is sufficiently
heavy that, already in the charm system, the KL correlation has become
less attractive than the JW correlation. The strict CM picture would thus
predict different structures for the $\theta$ and
its heavy quark analogues. This does {\it not} mean that the CM picture would
yield the JW mass estimates given in Eq. (\ref{jwthetac}). Indeed,
the JW correlation, which would dominate the heavy quark system,
produces only a portion of the hyperfine attraction in the
$\theta$ for CM interactions. Thus, in the CM model, a correction
for the reduction in the hyperfine expectation in the heavy quark
system would need to be added to the JW estimates. This correction to the
JW value moves the estimated $\theta_c$ mass to $\sim 20$ MeV above
the strong decay threshold. This effect, if present, would also
impact the estimates of Ref. [28] for the $P=-$,
$\bar{Q}s\ell^3$ states.

Interesting predictions of the minimally-model-dependent type
can be made for the $P=+$, $\bar{Q}\ell^4$ states in the GB and CM
models. The low-lying spin-flavor excitations for the two models
are shown in the table below. Numerical values for the splittings
from the ground-state pentaquark configuration have been
obtained by fully diagonalizing in the space of all Pauli-allowed,
fully antisymmetrized states for each channel, and using the
pair matrix elements from the baryon spectrum to estimate
the overall scale\cite{kmthetac}. One sees that a rather dense
spectrum of excitations is predicted, especially in the CM
model, and that the pattern of excitations is very different
for the two models. It also turns out
that the overlaps to the nucleon-plus-heavy-pseudoscalar decay 
channel are roughly comparable for all states listed 
(with the exception of one channel for the CM interactions
where the overlap is strongly suppressed). Since
the relative strengths of the couplings to the decay products
are expected to be given by the ratio of the corresponding 
overlap factors\cite{closezhao}, one expects a rich spectrum
of experimentally observable excited states.
Such predictions should be rather easy to confirm or rule out,
assuming any of the predicted states can be found experimentally.

\begin{table}[ph]
\tbl{Low-lying positive parity excitations of the $\theta_Q$ in the
GB and CM models, in the $m_{\bar{Q}}\rightarrow\infty$ limit.
$E_{ex}$ is the excitation energy in MeV.}
{\footnotesize
\begin{tabular}{@{}lrr@{}}
\hline
$\ (I,J_q)$&$E_{ex}$ (GB)&$E_{ex}$ (CM)\\
\hline
(0,1/2)&0&0\\
(0,1/2)&330&90 \\
(0,3/2)&330&90 \\
(1,1/2)&150&120\\
(1,1/2)&350&130\\
(1,3/2)&150&120\\
\hline
\end{tabular}\label{tablethetac} }
\vspace*{-13pt}
\end{table}

\section*{Acknowledgments}
The ongoing support of the Natural Sciences and Research
Engineering Council of Canada is gratefully acknowledged.
There are many topics and recent papers I have been unable
to discuss due to space restrictions. Apologies in advance
to the authors of those works.
%
%
%
%

\end{document}